\documentclass[letter]{aa} 

\usepackage{graphicx,natbib}
\bibpunct{(}{)}{;}{a}{}{,} 
\usepackage[varg]{txfonts}
\usepackage[colorlinks=true,citecolor=blue]{hyperref}
\usepackage[normalem]{ulem}
\usepackage{threeparttable}
\begin{document} 

   \title{Mind the gap: Distinguishing disc substructures and their impact on the inner disc composition}
   
   \author{Jingyi Mah\inst{1}
          \and
          Sofia Savvidou\inst{1}
          \and
          Bertram Bitsch\inst{2,1}
          }

   \institute{Max-Planck-Institut f\"{u}r Astronomie, K\"{o}nigstuhl 17, 69117 Heidelberg, Germany\\
             \email{mah@mpia.de; savvidou@mpia.de}
             \and
             Department of Physics, University College Cork, Cork, Ireland
             }

   \date{Received ; }

 
  \abstract
  {Improved observational technologies have enabled the resolution of substructures and the measurement of chemical abundances in protoplanetary discs. Understanding the chemical composition of the inner disc allows us to infer the building blocks available for planet formation. Recently, the depletion of water in the inner disc has been suggested to be linked to the presence of substructures, such as gaps and rings, further out in the disc. We investigate this hypothesis further by running 1D semi-analytical models of a protoplanetary disc with a gap to understand the combined effects of disc viscosity, gap depth, gap location, and gap formation timescales on the composition of the inner disc (water abundance, C/O, O/H, and C/H ratios). Our results show that for a specific value of disc viscosity, the simulation outcome can be classified into three regimes: shallow gap, 'traffic jam', and deep gap. While deep gaps may already be distinguishable with moderate-resolution (FWHM $\sim$ 10~AU) techniques, it is still challenging to resolve shallow gaps with the current capabilities. On the other hand, discs with traffic jams have a higher chance of being resolved when observed with a high resolution (FWHM $\lesssim$ 5~AU), but they may appear as an intensity enhancement or even featureless when observed with moderate to low angular resolution (FWHM $\gtrsim$ 10~AU). In this regard, information on the inner disc composition is useful because it can help to infer the existence of traffic jams or distinguish them from deep gaps: discs with deep gaps are expected to have a low water content -- and thus high C/O ratio in the inner disc due to the effective blocking of pebbles -- while discs with shallow gaps would demonstrate the opposite trend (water-rich and low C/O ratio). Furthermore, discs with a traffic jam would have a constant (albeit low) inward flux of water-rich pebbles resulting in a moderate water content and sub-stellar C/O ratios. Finally, we find that the effectiveness of gaps as pebble barriers diminishes quickly when they form late $(t_{\rm gap} \gtrsim 0.1~{\rm Myr})$, as most of the pebbles have already drifted inwards.}


   \titlerunning{How different disc substructures influence the composition of the inner disc}
   \authorrunning{J. Mah, S. Savvidou and B. Bitsch}
   
   \maketitle
%
\section{Introduction}
High angular-resolution observations of protoplanetary discs with the Atacama Large Millimetre/submillimetre Array (ALMA) reveal that substructures such as bright `rings’ and dark `gaps’ are a common occurrence for bright and large $(R_{\rm eff} \geq 50~{\rm AU})$ discs \citep[e.g.][]{Andrewsetal2018, Longetal2018, Ciezaetal2021}. Recent attempts at a re-analysis have uncovered the presence of substructures at smaller orbital distances as well as in discs previously thought to be featureless \citep{Jenningsetal2022,Zhangetal2023}, indicating that substructures may be a universal feature and that we may only be limited by the resolution of observations. Rings have been suggested to be dust traps where dust particles pile up and accumulate \citep{Dullemondetal2018}. This disrupts the otherwise smooth inward drift of dust and larger-size solids called `pebbles' towards the star, causing a depletion in the solid surface density interior to the dust trap.

There are multiple proposals to explain the origin of these disc substructures: pressure perturbations generated by hydrodynamical instabilities or zonal flows that form spontaneously in the disc \citep[e.g.][]{Flocketal2015}, the location of major ice lines (or condensation or evaporation fronts) where dust grains grow rapidly \citep{Zhangetal2015} and pile up due to an opacity transition at the ice lines \citep{Muelleretal2021}, or the presence of a growing planet that carves out a gap in the gas disc \citep[e.g.][]{PaardekooperMellema2006,Pinillaetal2012}. The latter proposal is perhaps the most appealing due to its broader implications, but it has been argued that this could not be true for all cases because having a planet in each gap of the same disc would lead to dynamical instabilities \citep[e.g.][]{Tzouvanouetal2023}. 

Despite the debate around the origin of disc substructures, it has become important to acknowledge and consider their effects in studies of planet formation as they could potentially change the outcome in a significant way. For example, the filtering and blocking of inward drifting pebbles by gaps reduce the amount of material that flows into the inner disc, which not only prevents planet growth \citep[e.g.][]{Izidoroetal2021}, but can also alter the chemical composition of the inner disc \citep{SchneiderBitsch2021a,Kalyaanetal2023}. Previous works have investigated the evolution of discs with gaps and introduced the gaps either in an ad hoc manner or self-consistently by modelling gap opening with growing planets \citep[e.g.][]{SchneiderBitsch2021a,Kalyaanetal2023}. However, the parameters examined in these studies are mostly related to the depth (or strength) of the gaps, while variations in the timing when the gap appears have not been explored in detail.

More recent observational data obtained with the {\it James Webb} Space Telescope (JWST) reveal that some discs are water-poor in the inner regions \citep[e.g.][]{Banzattietal2023,Taboneetal2023} while others are water-rich \citep[e.g.][]{Banzattietal2023,Gasmanetal2023,Perottietal2023}. The difference in the inner disc water abundance has been suggested to be related to the {trapping of icy pebbles in dust traps located in the outer disc} \citep{Banzattietal2023}. In an earlier study, \citet{BosmanBanzatti2019} showed that the depletion of carbon and oxygen by a factor of $\sim 50$ in the inner disc of TW Hya compared to the interstellar medium value is due to a dust trap that impedes the inward drift and evaporation of carbon and oxygen-rich ices.

The challenge now is to establish a link between observational data and modelling results. This is not straightforward because there are many parameters involved. Here, we present a compact study to investigate how the time of gap opening, in addition to the depth and location of the gap, influences the inward flux of pebbles and the corresponding variations in the chemical composition of the inner disc. Our aim is to understand what consequences the opening of gaps have on the chemical composition of inner discs, where we focus on the water and the C/H, O/H and C/O abundances. We then model the intensity profiles of discs with different gap properties to determine whether these gaps are observable with the technology currently available.

\section{Methods}
\label{sec:methods}
\subsection{Disc substructures}
To understand how different properties of the gap influence the composition of the inner disc, we modelled the time evolution of a protoplanetary disc with a gap using the 1D disc evolution code {\tt chemcomp} \citep{SchneiderBitsch2021a}. We used this code in the past to determine the disc composition for discs without substructures \citep{Mahetal2023} and showed how the disc's C/O ratio evolves with time for different stellar types. Here, we describe our implementation of the gap. The details of the gas disc evolution, dust growth, and chemical model are described in Appendix~\ref{appendix:disc_model}. 

We introduce a gap in the disc at a specific time, $t_{\rm gap}$, by modifying the profile of $\alpha$ to include a numerical factor that describes the profile of the gap \citep{Dullemondetal2018,SchneiderBitsch2021a}:
\begin{equation}
    \aleph(r) = 1 - [1-f_{\rm gap}]\exp\left(\frac{1}{2}\left[\frac{(r - r_{\rm gap})}{\sigma}\right]^2\right)~,
\end{equation}
where $0 < f_{\rm gap} \leq 1$ is the depth of the gap (lower values correspond to a deeper gap), $r_{\rm gap}$ is the location of the gap, and $\sigma$ is the width of the gap, expressed as:
\begin{equation}
    \sigma = 2 H_{\rm g}(r = r_{\rm gap})/[2\sqrt{2\log(2)}]~,
\end{equation}
with $H_{\rm g}(r = r_{\rm gap})$ as the gas scale height at the location of the gap in the unperturbed disc. The modified $\alpha$-viscosity profile is then:
\begin{equation}
    \alpha' = \alpha/\aleph(r)~,
\end{equation}
where $\alpha$ is a fixed value throughout the disc except at $r = r_{\rm gap}$. This modification of $\alpha$ to $\alpha'$ is only applied when we compute the time evolution of the gas surface density. The original value of $\alpha$ is used for other physical processes that involve $\alpha$ (e.g. pebble fragmentation). In this work, we explore the outcome of varying $\alpha$, in combination with variations in $r_{\rm gap}$, $f_{\rm gap}$ and $t_{\rm gap}$. The initial conditions of our simulations are summarised in Table~\ref{tab:initial_conditions}. 

\begin{table}
    \caption{Initial conditions of our simulations.}
    \label{tab:initial_conditions}
    \centering
    \begin{tabular}{lcc}
    \hline\hline
        Parameter & Symbol & Values\\
        \hline
        Stellar mass & $M_*$ & $1~M_{\odot}$ \\
        Stellar metallicity & {\rm [Fe/H]} & 0.0 \\
        Disc mass & $M_{\rm disc}$ & $0.1~M_{\odot}$ \\
        Disc radius & $R_{\rm c}$ & 100~AU \\
        Disc lifetime & $t_{\rm disc}$ & 10~Myr \\
        Disc metallicity & $\epsilon$ & 0.015 \\
        Disc viscosity & $\alpha$ & $3\times10^{-5}$, $10^{-4}$, \\
                       &  & $3\times10^{-4}$, $10^{-3}$ \\
        Fragmentation velocity & $u_{\rm frag}$ & 5~m\,s$^{-1}$ \\
        Gap location & $r_{\rm gap}$ & 3, 10~AU \\
        Gap depth & $f_{\rm gap}$ & 0.01, 0.05, 0.1, 0.2, \\
         & &  0.3, 0.4, 0.5, 0.6, \\
         & &  0.7, 0.8, 0.9, 1.0 \\
        Gap formation time & $t_{\rm gap}$ & 0, 0.1, 0.3, 0.5, 1.0~Myr \\
        \hline
    \end{tabular}
\end{table}

\subsection{Radial intensity profiles}
We discuss the observability of the gaps through the radial intensity profiles, normalised to the peak intensity along the radius of the disc. The intensity of the dust continuum emission is expressed as: 
\begin{equation}
    I_\lambda=B_\lambda(T)[1-\exp(-\tau_\lambda)]~,
\end{equation} 
where $B_\lambda(T)$ is the Planck spectrum at a temperature $T$ and $\tau_{\lambda}$ is the optical depth. In the Rayleigh-Jeans approximation, which is relevant for large wavelengths $\lambda$, the Planck spectrum is written
\begin{equation} 
    B_\lambda(T)=\frac{2 c k_{\rm B}T}{\lambda^4}~. 
\end{equation} 
Therefore, the intensity is:
\begin{equation} 
\label{eq:Intensity}
    I_\lambda=\frac{2 c k_{\rm B}\overline{T}}{\lambda^4}(1-\exp(-\overline{\kappa}\Sigma_Z))~,
\end{equation} 
where we adopt the standard assumptions of $\overline{T}$= 20 K \citep{AndrewsWilliams2005} and $\overline{\kappa}= 2.3~{\rm cm^2\,g^{-1}}$ (for $\lambda$ = 1.3mm) \citep{Beckwithetal1990}.
\section{Results}
\label{sec:results}
Here, we report the results for a disc with a gap at 10~AU which corresponds to a location just beyond the CO$_2$ ice line in our model. The results for a gap which is closer in at 3~AU are shown in Appendix~\ref{appendix:3au_gap}.

\subsection{Inner disc composition as a function of gap depth}

\begin{figure*}
\centering
   \resizebox{\hsize}{!}{\includegraphics{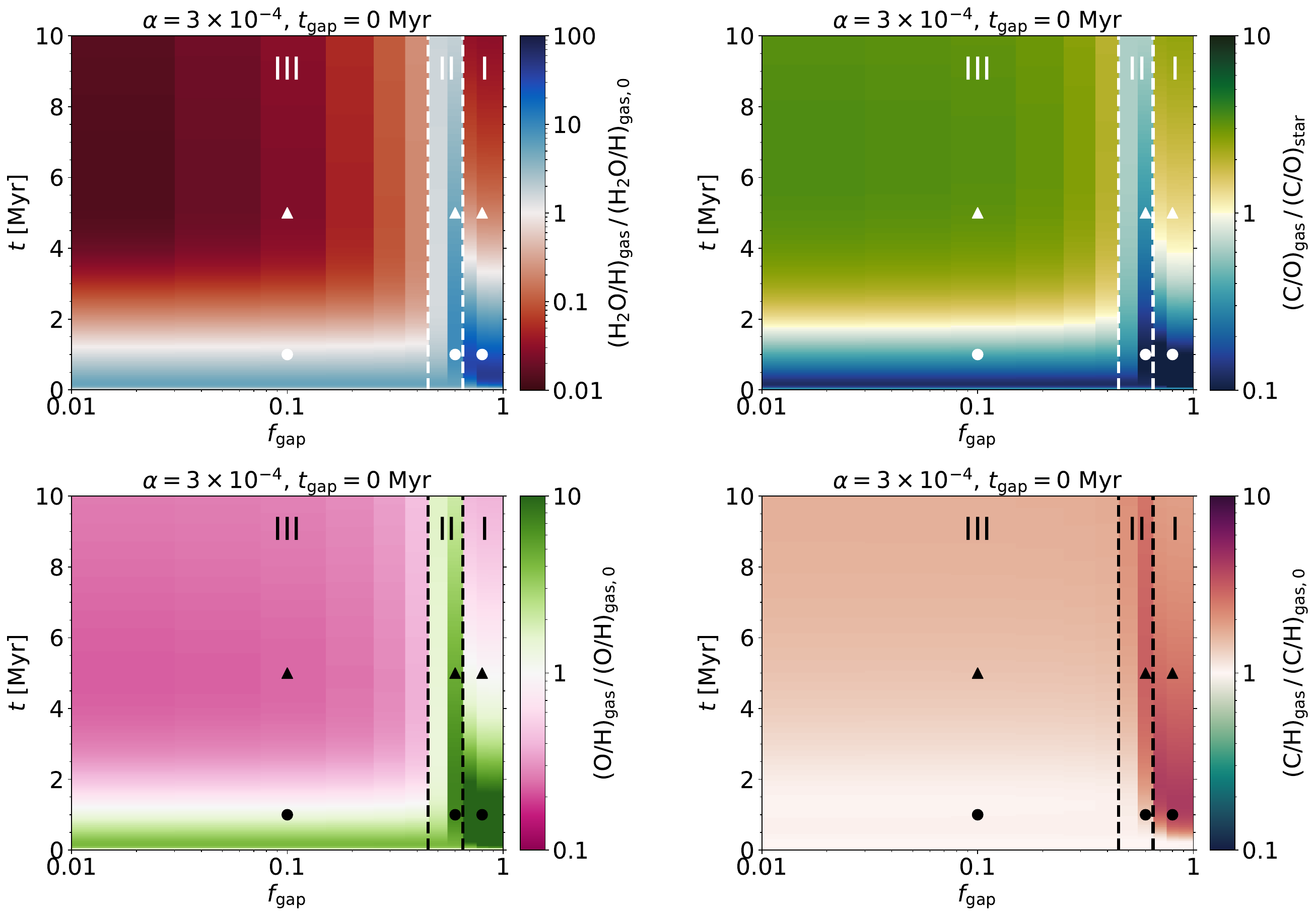}}
   \caption{Time evolution of the normalised water vapour abundance, C/O, O/H, and C/H abundance ratios at $r = 0.5~{\rm AU}$ as a function of gap depth for a disc with a gap at 10~AU. We plot the results for disc viscosity $\alpha = 3\times10^{-4}$ and pebble fragmentation velocity $u_{\rm frag} = 5~{\rm ms}^{-1}$. Regimes I, II, and III correspond respectively to the scenario of shallow gap, traffic jam, and deep gap. Circles and triangles are selected examples from the three regimes at different time snapshots where we further investigate their observability (see Section~\ref{sec:results_observability}).}
   \label{fig:bump-10au_3e-4}
\end{figure*}

In Fig.~\ref{fig:bump-10au_3e-4}, we show a compilation of the time evolution of various measurable chemical abundances (water vapour abundance, the C/O, O/H, and C/H ratios) as a function of gap depth, $f_{\rm gap}$, measured at $r = 0.5~{\rm AU}$ for a disc with fixed disc viscosity, $\alpha = 3\times10^{-4}$, and gap formation time, $t_{\rm gap} = 0~{\rm Myr}$. The C/O ratio is normalised to the stellar value (0.55) while the abundance of water, oxygen, and carbon are normalised to the inner disc's value at $t=0$ and $r=0.5~{\rm AU}$ ((H$_2$O/H)$_{\rm gas,0} = 3.54\times10^{-4}$, (O/H)$_{\rm gas,0} = 4.91\times10^{-4}$, (C/H)$_{\rm gas,0} = 1.37\times10^{-4}$). Our simulation results can be roughly divided into three regimes, which we labelled as I, II, and III in the figure.

Regime I corresponds to the scenario of a shallow gap which is ineffective at blocking the inward drift of pebbles. The inner disc is initially enriched in water vapour due to the evaporation of water ice-rich pebbles. This is also reflected in the increase in the O/H ratio and a decrease in the C/O ratio to sub-stellar values due to the increase in the abundance of oxygen compared to carbon. The duration of this initial phase depends on the disc’s viscosity and its mass, where the former determines the speed of gas transport and the latter sets the total amount of solids available. In this example with $\alpha = 3\times10^{-4}$, the water vapour enrichment in the inner disc lasts for about 3~Myr. Over time, the oxygen abundance will dwindle down due to the loss of both water vapour and pebbles to the star. The inner disc then becomes depleted in oxygen and enriched in carbon due to the inward advection of CO$_2$ gas from the evaporation of pebbles initially located interior to the gap. This causes the C/O ratio in the inner disc to rise and reach super-stellar values. This result is in agreement with our previous study \citep{Mahetal2023} where no gaps were included.

Regime II represents the scenario of a `traffic jam' whereby the gap is just deep enough to block a fraction of the pebble flux but still allows a steady (but reduced) flux of small pebbles to enter the inner disc. The evaporation of the stream of small pebbles in the inner disc provides a low but long-lived enrichment of water vapour that can last for more than 8~Myr in our example. Due to the longevity of the oxygen enrichment, the C/O ratio remains sub-stellar even though carbon-rich gas is slowly transported into the inner disc.

Regime III reflects the scenario of a deep gap capable of obstructing the flow of inward drifting pebbles. The initial enrichment of water vapour in the inner disc is low and it does not last long (about 1~Myr in our example), as it is set by the water-rich pebbles available initially interior to the gap. The C/O ratio does not jump to super-stellar values before $t = 2~{\rm Myr.}$ This is because, in addition to blocking water ice rich pebbles from reaching the inner disc, the deep gap at 10~AU also halts pebbles that contain CO$_2$ just beyond the CO$_2$ ice line, preventing them from evaporating and enriching the gas. The evolution of the C/O ratio in the inner disc depends solely on the disc's viscosity, which determines how long water-rich gas can remain and how long it takes for carbon-rich gas to arrive. 

\subsection{Observability of the gaps}
\label{sec:results_observability}
\begin{figure*}
\centering
   \resizebox{\hsize}{!}{\includegraphics{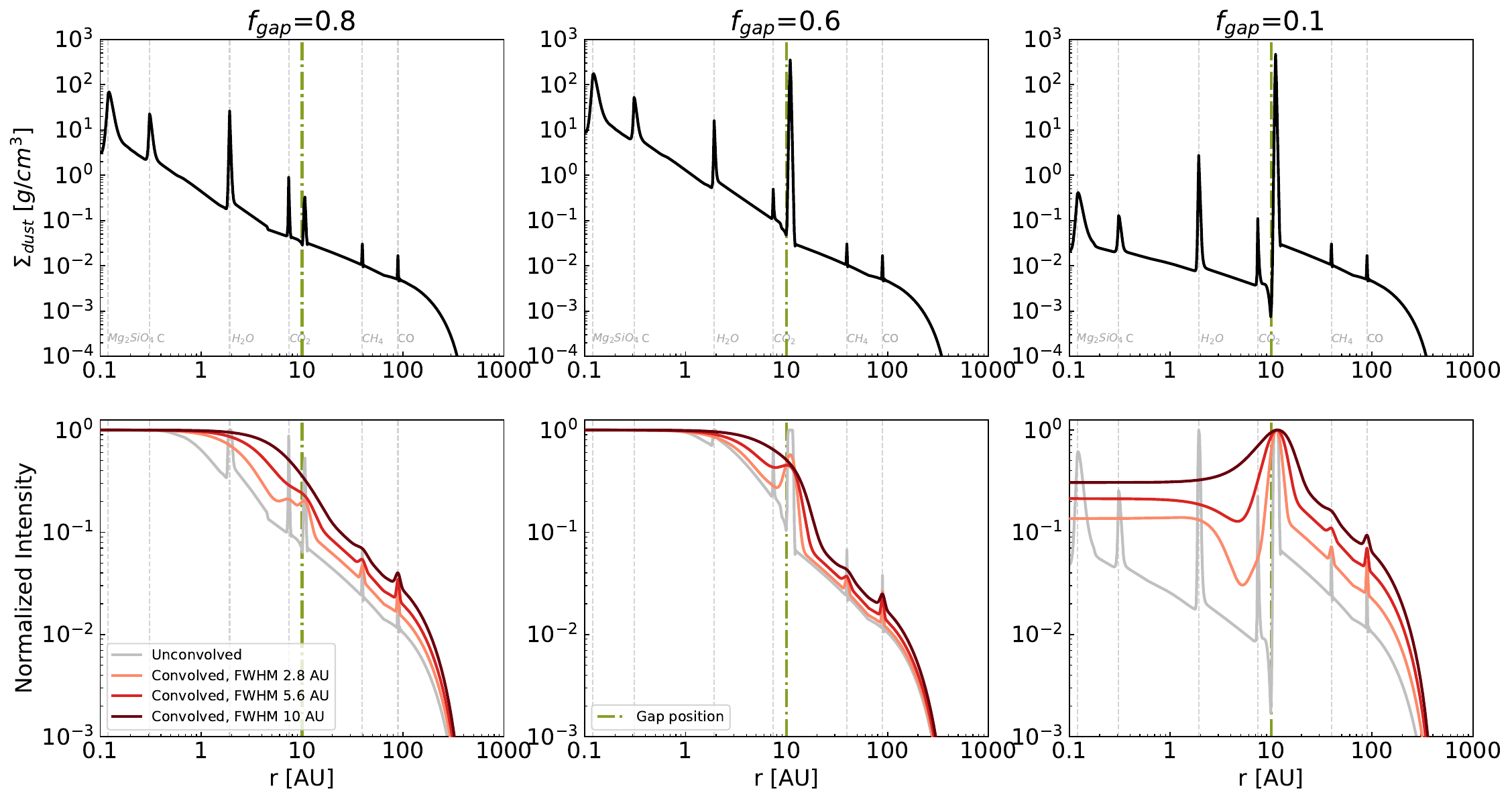}}
   \caption{ Surface density of solids (top) and normalised intensity (bottom) as a function of orbital distance at 1 Myr for three gap depths (marked by circles in Fig.~\ref{fig:bump-10au_3e-4}). We compare the unconvolved intensity with the ones convolved with three different beams. The gray dashed lines show the evaporation fronts that mainly cause the spikes in the intensity.}
\label{fig:intensity_plot_1Myr}
\end{figure*}

The regimes that were discussed in the previous section lead to different features in the disc that may or may not be observable. The strength of the spike in the surface density of the solids depends on the depth of the gap (top row plots in Fig.~\ref{fig:intensity_plot_1Myr}). The different gap depths shown in this plot (0.8, 0.6, and 0.1) correspond to the three aforementioned regimes (I, II, and III) at $t = 1~{\rm Myr}$ and the circles shown in Fig.~\ref{fig:bump-10au_3e-4}. The rest of the spikes in the surface density are the pile-ups caused by the evaporation of the chemical species at their corresponding ice lines (gray dotted lines). We note that the efficient trapping of pebbles at the gap reduces the pebble surface density in the inner disc and, consequently, results in less efficient planet formation in the inner disc. This is in line with the simulations of \citet{Izidoroetal2021} which show that an early blocking of pebbles is necessary to form the terrestrial planets instead of super-Earths.

We can also calculate, using Eq.~\ref{eq:Intensity}, the radial intensity profiles normalised to the peak intensity along the radius of the disc. We show the normalised radial intensity convolved with three distinct beams (bottom row plots in Fig.~\ref{fig:intensity_plot_1Myr}). The full width at half maximum (FWHM) of the first two beams is 0.02" and 0.04", equivalent to 2.8 and 5.6~AU, respectively, assuming a source located at an average distance of 140~pc. These selections represent some of the most detailed observations to date \citep[e.g.][]{Andrewsetal2018,Benistyetal2021}. We also included a convolution with a beam where the FWHM would be 10 AU, considering this as a lower limit for moderate resolution. 

The deepest gap (regime III, $f_{\rm gap} = 0.1$) would probably be recovered even with moderate resolution. The obstruction of the pebble flow towards the inner disc creates a high contrast in the brightness between the ring and the gap that precedes it, which makes it easier to identify as a strong feature in the disc. The observability of the shallowest gap (regime I, $f_{\rm gap} = 0.8$) depends not only on very high angular resolution but also potentially on a source that is located in a nearby region. The gap in the intermediate case of regime II ($f_{\rm gap} = 0.6$) would be easily detected if observed with an angular resolution at the highest end of our current limits. If the resolution corresponds to a FWHM of $\sim 2.8~{\rm AU,}$ then there is enough contrast to identify the substructure. However, moving towards moderate resolution (FWHM $\geq$ 10 AU) indicates that the feature would more likely be considered an enhancement to the underlying surface density profile. In Appendix~\ref{appendix:5Myr_intensity_profile}, we show the surface density and intensity profile at 5~Myr (corresponding to the triangles in Fig.~\ref{fig:bump-10au_3e-4}).

\subsection{Importance of early gap formation}
\label{results:gap_formation_time}

We show how the water abundance, C/O, O/H, and C/H ratios are influenced by the gap formation timescales in Appendix~\ref{appendix:additional_figures}. 

At low viscosities ($\alpha \lesssim 10^{-4}$), our simulations show that the the inner disc is water-rich for 10~Myr in the presence of a shallow gap ($f_{\rm gap} \gtrsim 0.8$). In this simulation setup, the pebbles are large and thus release more water vapour to the gas per unit time. However, they also drift quickly towards the star. On the other hand, the long viscous timescale of the disc allows the water vapour to linger around in the inner disc for a long period, resulting in a net long-lived water vapour enrichment and sub-stellar C/O ratios. 

In the presence of a deep gap ($f_{\rm gap} \lesssim 0.7$), most of the pebbles are blocked and the abundance of water vapour in the inner disc is indeed reduced. Despite this, the water vapour enrichment in the inner disc continues for a protracted period of time due to the low disc viscosity. Carbon-rich gas also takes a long time to arrive at the inner disc. Consequently, the C/O ratio remains at sub-stellar values for more than 5~Myr. Our results also show that the parameter space for the traffic jam scenario is narrow, suggesting that this outcome could be rather rare for low-viscosity discs.
 
At higher disc viscosities ($\alpha \gtrsim 3\times10^{-4}$), the duration of water vapour enrichment is shorter for all three regimes. This is mainly contributed by the faster accretion of gas onto the central star and the smaller amount of water vapour released to the gas by smaller sized pebbles. We also see that the parameter space for the traffic jam scenario broadens. For the case of $\alpha = 10^{-3}$, there is no value of $f_{\rm gap}$ below which we are in the deep gap scenario (regime III) within the range of values for $f_{\rm gap}$ that we explored. The traffic jam regime extends from $f_{\rm gap} \approx 0.1$ down to at least $f_{\rm gap} \approx 0.01$ because dust sizes are so small that they pass through the gap easily with the gap just acting to regulate the flow of the pebbles. Since the disc viscosity is high, carbon-rich gas from the outer disc is advected inwards quickly. However, due to the lower abundance of carbon compared to oxygen, the overall C/O ratio remains at sub-stellar values.

When we delay the introduction of the gap in the disc (moving from left to right in the figures), we clearly see an increase in the duration and level of water vapour enrichment for regime III (deep gap; lower $f_{\rm gap}$) because now the water ice-rich pebbles arrive in the inner disc before the gap forms. Even if we introduce a deep gap slightly later at $t_{\rm gap} = 0.1~{\rm Myr}$, the inner disc can still be enriched in water vapour for at least 8~Myr in a low-viscosity disc ($\alpha \leq 10^{-4}$) and up to 2~Myr if $\alpha \geq 3\times10^{-4}$, demonstrating the strong influence of pebble drift and evaporation on the water abundance in the inner disc. Following the same trend, the traffic jam effect in regime II (region between the white dashed lines) diminishes with later gap formation times. As expected, the outcomes for both deep gap and traffic jam scenarios approach that of a shallow gap as we introduce the gap at increasingly later times. Our results demonstrate the importance of an early gap formation time in countering the effect of pebble drift and reducing the abundance of water vapour in the inner disc.

\section{Discussion and summary}
\label{sec:discussion}

Through our exploration of the consequences of the range of gap depths on the elemental abundances in the inner disc, we show that there exist three possible regimes in terms of gap depth -- deep gap, shallow gap, and traffic jam -- and they could potentially be distinguished with current observational capabilities. However, the detection of substructures in discs requires spatially resolved emission and is thus heavily affected by observational as well as image reconstruction biases. A lack of substructures in observed discs could be correlated with insufficient angular resolution \citep{Baeetal2023}. At the same time, the chosen deconvolution technique could lead to loss of information \citep{Jenningsetal2022}. In many cases, the substructures have more chances of being detected if one were to fit for the visibilities, rather than the reconstructed images \citep{Pearson1999}. 

Nevertheless, the combination of observed substructures as well as measurements of various elemental abundances in the inner disc could take us one step forward in distinguishing between deep gaps and traffic jams and constraining important properties (such as the viscosity) of the disc. For example, acquiring compositional data for the inner discs of the DSHARP sample \citep{Andrewsetal2018} would allow us to classify the substructures with more confidence. For discs observed at moderate to low resolution, compositional data would help us identify sources with substructures (which have not yet been detected) for future surveys at a higher resolution.

In \citet{Kalyaanetal2023}, the authors concluded that the gap that is closest to the water ice line plays the most important role in determining the abundance of water vapour in the inner disc. This is because the efficiency of the gap also depends on how much of the solids it can block. In the case of a gap at 10~AU, dust grains initially located interior to the gap can drift inwards and evaporate, thereby contributing to the water vapour enrichment in the inner disc. Our simulations with a gap at 3~AU (Appendix~\ref{appendix:3au_gap}) confirm that the gap location indeed results in a lower abundance of water in the inner disc compared to the case of a gap at 10~AU. As it is still challenging to detect substructures very close to the star, information on the inner disc's composition can help to suggest the presence or absence of substructures. 

Our analysis further reveals the intricate relationship between disc viscosity and gap properties, which work together to influence the composition of the inner disc. The presence of a deep gap is no doubt important to halt drifting pebbles, but the timing when the gap appears in the disc is equally crucial in maximising its effectiveness; however, this parameter has not yet been extensively studied. Due to the swift inward drift of pebbles, we find gaps should be established almost immediately or within the first 0.1~Myr of the disc's evolution for them to be effective.

\citet{Banzattietal2023} used JWST to determine the water content of four inner discs: two are compact discs with no visible disc substructures\footnote{The beam resolution is around 10-15 AU, making it hard to detect substructures in the first place} and two are larger discs with visible gaps. They found that the small discs contain water vapour in the inner regions, while the large discs are depleted in water vapour in the inner regions. Our study suggests that the small compact discs investigated in that paper (GK Tau and HP Tau) would have only minimal substructures. On the other hand, the substructures of the large discs must have formed early in order to avoid a large enrichment of the inner disc with water vapour. The difference in our interpretation of the data compared to \citet{Banzattietal2023} is two-fold: first we attribute the difference in the water abundance not to the disc size, but to substructures that are able to hinder (or even block) the pebble flux; secondly, our model suggests a time component for when the substructures appeared.

In the PDS 70 system, where a wide gap is present in the outer disc with two giant planets located in the gap \citep{Keppleretal2018,Muelleretal2018}, water molecules have been identified in the inner disc \citep{Perottietal2023}. The longevity of the inner disc is intriguing as it is expected that the gap carved by the pair of giant planets would be deep. Recent modelling results show that only sub-micrometre-size dust grains entrenched in the gas can seep into the gap \citep{Pinillaetal2024}. This example would correspond to the traffic jam scenario we present in this work where the water vapour in the inner disc can last for at least a few Myr if the disc viscosity is $\alpha \leq 3\times 10^{-4}$.

It is then possible to then follow up and consider whether is it possible to form a deep gap early in the disc. Assuming that the gap is carved by growing planets, in a disc of $\alpha = 3\times10^{-4}$, a gap of depth $f_{\rm gap} = 0.6$ at 10~AU capable of generating a traffic jam would correspond to a planetary core with a planet-to-star mass ratio of $q \approx 2\times10^{-5}$ (see Appendix~\ref{appendix:fgap_vs_q}). This is equivalent to a $6.5~M_{\oplus}$ core if the mass of the central star is $1~M_{\odot}$. The latest numerical simulations show that planetary embryos around a $1~M_{\odot}$ star take about 0.1 to 0.15~Myr to reach pebble isolation mass \citep{SavvidouBitsch2023}. In the inner disc, the growth timescale of planetary embryos is at least 0.1~Myr under favourable conditions \citep{Voelkeletal2021a,Voelkeletal2021b} while it will take longer ($\sim 0.5~{\rm Myr}$) at larger orbital distances. Taken together, massive planets which could carve a deep gap in the disc could be challenging to form within 0.1~Myr, unless the disc is very massive to begin with.

We have also made the assumption that the pebbles have a maximum size corresponding to fragmentation velocity of 5~${\rm m\,s^{-1}}$. Recent observations suggest that icy pebbles are fragile and could possibly be much smaller than what we assume here \citep{MusiolikWurm2019,Jiangetal2024}. If this is indeed the case, it would imply that the three regimes we have identified might be shifted to deeper gaps, which are more efficient at blocking small particles \citep[e.g.][]{Ataieeetal2018,Bitschetal2018}.

In the interest of simplicity, we did not take the effect of planetesimal formation into account in this work. The formation of planetesimals takes away planetary building blocks, thereby reducing the enrichment due to evaporating pebbles \citep[e.g.][]{Dantietal2023}. This would result in an overall reduction in the abundances we see in our simulations. \citet{Kalyaanetal2023} reported that the influence of planetesimal formation on the vapour enrichment in the inner disc is weak when a gap is already present in the disc because the planetesimals would form at the pressure bump exterior to the gap where the local dust-to-gas mass ratio is sufficiently high; thus, it would not be able to influence the inner disc regions.

To summarise, we have set out to link observational constraints with theory by studying how the composition of the inner disc changes in the presence of a gap further out in the disc. The first aim of our work is to understand how the interplay between disc viscosity and the various properties of the gap (depth, location, and formation time) influence the volatile abundances and C/O ratio of the inner disc. The second aim is to determine whether gaps of different depths which offer discernible outcomes in our simulations can be sufficiently resolved with the current observational capabilities.

We show that the gaps can be classified into three regimes, while the question of whether or not they can be resolved depends on the contrast and angular resolution of the observations:
\begin{itemize}
    \item Shallow gap ($f_{\rm gap} \gtrsim 0.8$, regime I in Fig.~\ref{fig:bump-10au_3e-4}): presents a weak barrier that is inefficient at blocking inward drifting pebbles. It is challenging to observe unless the resolution is high and the source is nearby.
    \item Traffic jam ($0.5 \lesssim f_{\rm gap} \lesssim 0.6$, regime II in Fig.~\ref{fig:bump-10au_3e-4}): intermediate case where most pebbles are blocked, but some dust grains can flow through at a reduced rate. These would most likely be smeared out by the beam convolution at moderate resolution, so a higher resolution is required to resolve them. However, even enhancements with a low contrast could be identified as traffic jams if combined with compositional data. Similarly, traffic jams can also appear as rings and additional information on the composition of the inner disc becomes helpful to distinguish them from rings associated with deep gaps. These types of substructures could be caused by disc turbulence or growing planets that have yet to reach a pebble isolation mass (Appendix~\ref{appendix:fgap_vs_q}).
    \item Deep gap ($f_{\rm gap} \lesssim 0.4$, regime III in Fig.~\ref{fig:bump-10au_3e-4}): presents a strong barrier to block inwardly drifting pebbles. Observable even with moderate resolution (FWHM $\sim$ 10~AU). These could be caused by disc turbulence or, if $\alpha \leq 3\times10^{-4}$, gap-opening planets (Appendix~\ref{appendix:fgap_vs_q}).
\end{itemize}

Moving forward, we envision that new observational constraints on the composition of the inner disc will be helpful to distinguish whether the bright rings in ALMA observations are deep gaps or traffic jams, and to potentially identify substructures in discs that appear (almost) smooth in low to moderate resolution: a low water content and high C/O ratio would suggest that the disc has a deep gap. This prevents the inward drift of water-rich pebbles, while a moderate water content and sub-stellar C/O ratios would suggest the presence of a traffic jam that allows a constant, but a low inward flux of water-rich pebbles.

\begin{acknowledgements}
      We thank the anonymous reviewer for the insightful comments which helped to improve the clarity of this Letter. The authors acknowledge the support of the DFG priority program SPP~1992 ``Exploring the Diversity of Extrasolar Planets'' (BI~1880/3-1).
\end{acknowledgements}

\bibliographystyle{aa} 
\bibliography{50322corr} 

\begin{appendix}
\section{Disc model}
\label{appendix:disc_model}

\begin{table}
\centering
    \caption{List of elements and their abundances from \citet{Asplundetal2009} included in our model.}
    \label{tab:comp_sun}
    \begin{tabular}{c c}
    \hline\hline
    Element & Abundance\\ \hline
    He/H & 0.085 \\
    O/H  & $4.90\times10^{-4}$ \\
    C/H  & $2.69\times10^{-4}$ \\    
    Mg/H & $3.98\times10^{-5}$ \\
    Si/H & $3.24\times10^{-5}$ \\
    Fe/H & $3.16\times10^{-5}$ \\ \hline
\end{tabular}
\end{table}

\begin{table}
\centering
    \caption{List of chemical species included in our model.}
    \label{tab:chempartition}
    \begin{tabular}{c c c}
    \hline\hline
    Species (Y) & $T_{\text{cond}}$~[K] & Volume mixing ratio \\ 
    \hline
    CO              & 20   & 0.20 $\times$ C/H \\
    CH$_4$          & 30   & 0.10 $\times$ C/H \\
    CO$_2$          & 70   & 0.10 $\times$ C/H \\
    H$_2$O          & 150  & O/H - (CO/H + 2 $\times$ CO$_2$/H + \\
                    &      & 4 $\times$ Mg$_2$SiO$_4$/H + 3 $\times$ MgSiO$_3$/H) \\ 
    C               & 631  & 0.60 $\times$ C/H \\
    Mg$_2$SiO$_4$   & 1354 & Mg/H - Si/H \\
    Fe              & 1357 & Fe/H \\
    MgSiO$_3$       & 1500 & Mg/H - 2 $\times$ (Mg/H - Si/H) \\
    \hline
    \end{tabular}
\end{table} 

\subsection{Gas evolution}
We consider a gas disc with surface density profile described by the self-similar solution \citep{Lynden-BellPringle1974,Lodatoetal2017}
\begin{equation}
    \Sigma_{\rm g}(r,t) = \frac{M_{\rm disc}}{2\pi R_{\rm c}^2}(2-\psi)\left(\frac{r}{R_{\rm c}}\right)^{-\psi}\xi^{\frac{5/2-\psi}{2-\psi}}\exp\left(-\frac{(r/R_{\rm c})^{2-\psi}}{\xi} \right)~,
\end{equation}
where $M_{\rm disc}=0.1~M_{\odot}$ is the initial disc mass, $R_{\rm c}=100~{\rm AU}$ is the disc's characteristic radius, $\psi = ({\rm{d}} \ln \nu / {\rm{d}} \ln r) \approx 1.08$ is the logarithmic gradient of the viscosity evaluated at the disc inner edge, and the normed time $\xi = 1+t/t_{\nu}$, with the viscous time $t_{\nu}$ given by 
\begin{equation}
    t_{\nu} = \frac{R_{\rm c}^2}{3(2-\psi)^2 \nu(r = R_{\rm c})}~.
\end{equation}

The disc evolves in time according to the viscous advection-diffusion equation
\begin{equation}
    \frac{\partial{\Sigma_{\rm{g,Y}}}}{\partial t} - \frac{3}{r}\frac{\partial}{\partial r}\left[ \sqrt{r}\frac{\partial}{\partial r} \left( \sqrt{r}\nu \Sigma_{\rm{g,Y}} \right) \right] = \Dot{\Sigma}_{\rm Y}~,
\end{equation}
where $\nu$ is the disc’s kinematic viscosity, and $\Dot{\Sigma}_{\rm Y}$ is the source term for a given chemical species Y which is computed by considering the effects of pebble evaporation and condensation at the ice lines (we refer the interested reader to \citealp{SchneiderBitsch2021a} for more details). The disc viscosity is expressed as \citep{Pringle1981}
\begin{equation}
    \nu = \alpha c_{\rm s}^2 \Omega_{\rm K}^{-1}~,
\end{equation}
where $\alpha$ is a dimensionless parameter that describes the strength of turbulence \citep{ShakuraSunyaev1973}, $\Omega_{\rm K} = \sqrt{GM_*/r}$ is the Keplerian angular velocity, and $c_{\rm s}$ is the isothermal sound speed which is in turn related to the disc’s midplane temperature $T_{\rm mid}$ following:
\begin{equation}
    c_{\rm s} = \sqrt{\frac{k_{\rm B}T_{\rm mid}}{\mu m_{\rm p}}}~,
\end{equation}
where $k_{\rm B}$ is the Boltzmann constant, $m_{\rm p}$ is the proton mass, and $\mu$ is the mean molecular weight. At the end of the disc lifetime (10~Myr in this work), the disc is assumed to dissipate away with a decay timescale of 10~kyr. 

The temperature profile of the gas disc is computed by considering the heat from viscous accretion and stellar irradiation. The heat from the stellar irradiation is related to the stellar luminosity via $Q_{\rm irr} = \phi L_*/(4\pi r^2)$ where $\phi = 0.05$ is assumed. The heat from viscous accretion is given by $Q_+ = \frac{9}{4}\Sigma_{\rm g} \nu \Omega_{\rm K}^2$. The midplane temperature is computed as \citep{SchneiderBitsch2021a}
\begin{equation}
    T_{\rm mid}^4 = T_{\rm visc}^4 + T_{\rm eff}^4 =\frac{3}{8}\frac{\tau_{\rm d}Q_+}{\sigma_{\rm SB}}+\frac{Q_{\rm irr}}{2\sigma_{\rm SB}}~,
\end{equation} 
where $\sigma_{\rm SB}$ is the Stefan-Boltzmann constant and $\tau_{\rm d}$ is the optical depth. For simplicity, the temperature profile is fixed with time. Our assumption is valid for solar-type stars which luminosity do not change much over time and thus can be approximated to be constant.

\subsection{Dust evolution}
In the disc, micrometre-sized dust grains coagulate and grow into pebbles with sizes limited by radial drift, fragmentation, and drift-induced fragmentation, based on the two-population model of \citet{Birnstieletal2012}. The initial amount of solids in the disc is $\epsilon = \Sigma_{\rm Z}/\Sigma_{\rm g} = 0.015$, where $\Sigma_{\rm Z}$ is the dust surface density. A fraction of the dust is assumed to be in the pebbles according to $\Sigma_{\rm peb} = f_{\rm m} \times \Sigma_{\rm Z}$, with $f_{\rm m} = 0.75$ in the fragmentation-limited regime and $f_{\rm m} = 0.95$ in the drift-limited regime. The size of a dust particle in each of the aforementioned regimes are \citep{Birnstieletal2012}
\begin{align}
    a_{\rm frag} &\simeq \frac{2}{3\pi}\frac{\Sigma_{\rm g}}{\alpha \rho_{\rm Z}}\frac{u_{\rm frag}^2}{c_{\rm s}^2}~,\\
    a_{\rm drift} &\simeq \frac{2}{\pi}\frac{\Sigma_{\rm Z}}{\rho_{\rm Z}}\frac{V_{\rm K}^2}{c_{\rm s}^2}\left(\frac{{\rm{d}} \ln P}{{\rm{d}} \ln r }\right)^{-1}~,
\end{align}
where $\rho_{\rm Z}$ is the dust particle density, $u_{\rm frag} = 5~{\rm m\,s}^{-1}$ is the fragmentation velocity, $V_{\rm K} = \Omega_{\rm K}r$ is the Keplerian velocity, and $({\rm {d}} \ln P/{\rm {d}} \ln r)$ is the gas pressure gradient. The time evolution of the dust surface density follows \citep{Birnstieletal2012,SchneiderBitsch2021a}
\begin{equation}
    \frac{\partial \Sigma_{\rm Z,Y}}{\partial t} + \frac{1}{r}\frac{\partial}{\partial r} \left[ r \left(\Sigma_{\rm Z,Y}\cdot\Bar{u}_{\rm Z}-\frac{\partial}{\partial r} \left( \frac{\Sigma_{\rm Z,Y}}{\Sigma_{\rm g}}\right) \cdot \Sigma_{\rm g} \nu \right) \right] = -\Dot{\Sigma}_{\rm Y}~,
\end{equation}
where $\bar{u}_{\rm Z}$ is the mass-weighted dust velocity. 

Our model further includes the effect of pebble evaporation and recondensation around the location of the ice lines \citep{SchneiderBitsch2021a}. This cycle enriches the gas interior to the ice lines with the vapour of the corresponding chemical species and generates a localised pile up of pebbles exterior to the ice lines.

\subsection{Chemical model}
We follow the chemical model of \citet{MahBitsch2023} which is a simplified version (omitting nitrogen and sulphur, as well as the majority of the refractory elements) of the original presented in \citet{SchneiderBitsch2021a}. This allows us to focus on the abundance of water and the major volatiles in the inner disc which contain carbon and oxygen. The elements included in our model can be found in Table~\ref{tab:comp_sun} and the partitioning model we employed to compute the initial disc composition is shown in Table~\ref{tab:chempartition}.


\section{Gap at 3 AU}
\label{appendix:3au_gap}
\begin{figure*}
\centering
   \resizebox{\hsize}{!}{\includegraphics{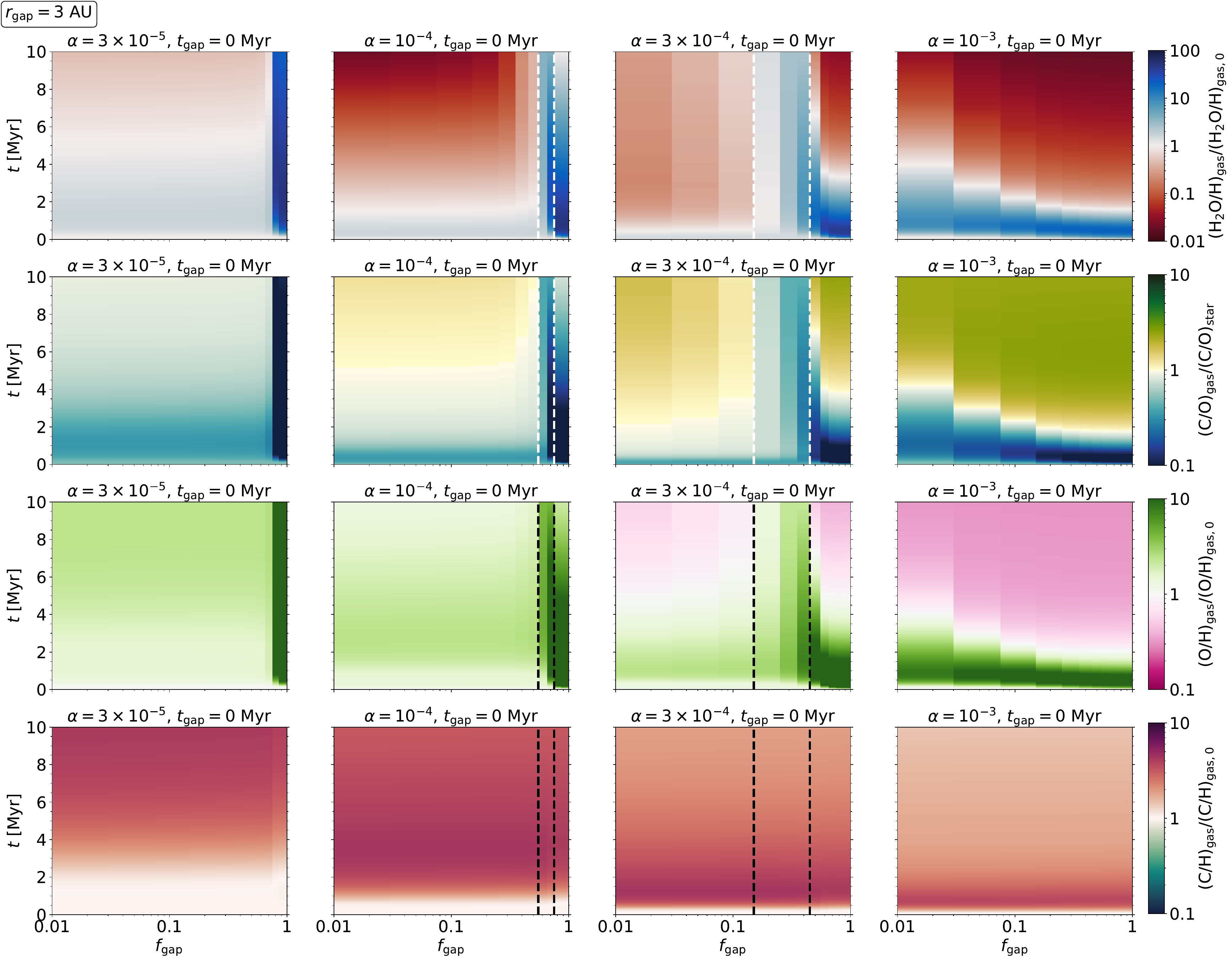}}
   \caption{Time evolution of the water abundance and abundance ratios of selected elements at 0.5~AU as a function of $\alpha$. Results are normalised to the initial values at this location. Here the gap is located at 3~AU (beyond the water ice line but interior to the CO$_2$ ice line) and forms immediately at $t_{\rm gap} = 0~{\rm Myr}$.}
   \label{fig:bump-3au}
\end{figure*}

When the gap is located at 3~AU, it can block more solids from entering the inner disc compared to a gap further out at 10~AU. As a consequence, the abundance of water vapour, which is strongly correlated to the O/H ratio, within the water ice line is strongly reduced (Fig.~\ref{fig:bump-3au}), especially at early times. The C/H ratio increases when carbon-rich gas contributed from the evaporation of CO$_2$ is advected inwards. This can take $\sim 0.5$~Myr to $\gtrsim 2$~Myr depending on the disc viscosity. 

\section{Disc intensity profile at 5 Myr}
\label{appendix:5Myr_intensity_profile}
\begin{figure*}
\centering
   \resizebox{\hsize}{!}{\includegraphics{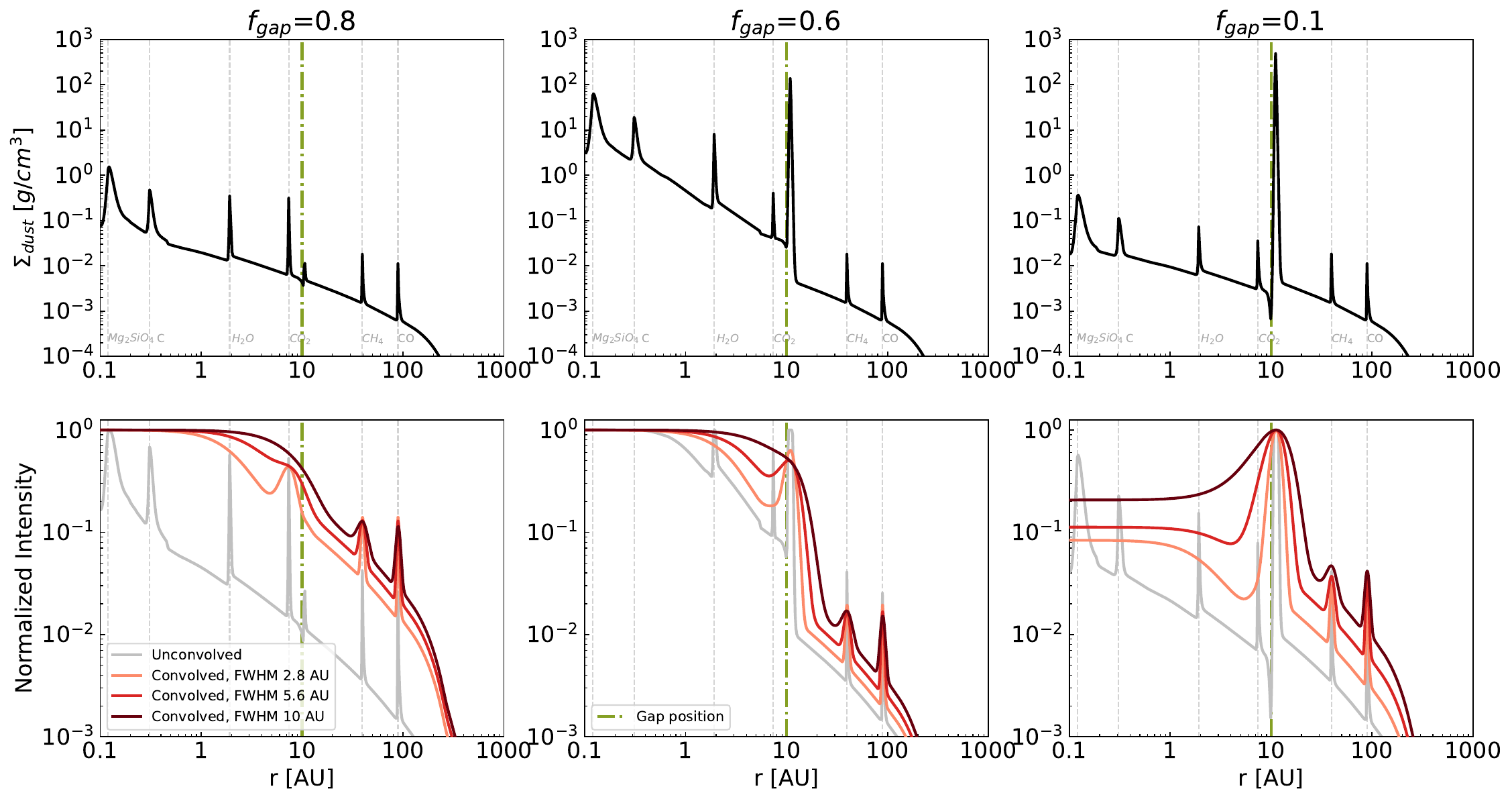}}
   \caption{Same as Fig. \ref{fig:intensity_plot_1Myr} at 5 Myr.}
   \label{fig:intensity_plot_5Myr}
\end{figure*}

As the disc evolves, the features become more prominent (Fig. \ref{fig:intensity_plot_5Myr}). In regime I ($f_{\rm gap}$ = 0.8), the ineffectiveness of the shallow gap to block the drifting pebbles leads to a significant decrease in the surface density of the dust over time. This also means that the feature in the normalized intensity because of the gap is not present anymore. Note that the bump in the intensity close to the location of the gap (only for the highest resolution) is instead caused by a spike in the dust surface density at the CO$_2$ ice line. Moving to regime II ($f_{\rm gap}$=0.6), we find that the time evolution and the `traffic jam' that the gap causes, makes for a clear `ring' if observed at high resolution, however it remains completely undetected with moderate resolution. Finally, the deep gap in regime III has effectively blocked the drifting pebbles exterior to its orbital location and therefore the bump in the intensity profile is larger at 5 Myr and the brightness contrast would be large enough to clearly detect a `ring' even with moderate resolution. 

\section{Evolution of inner disc composition as a function of gap formation time}
\label{appendix:additional_figures}
\begin{figure*}
\centering
   \resizebox{\hsize}{!}{\includegraphics{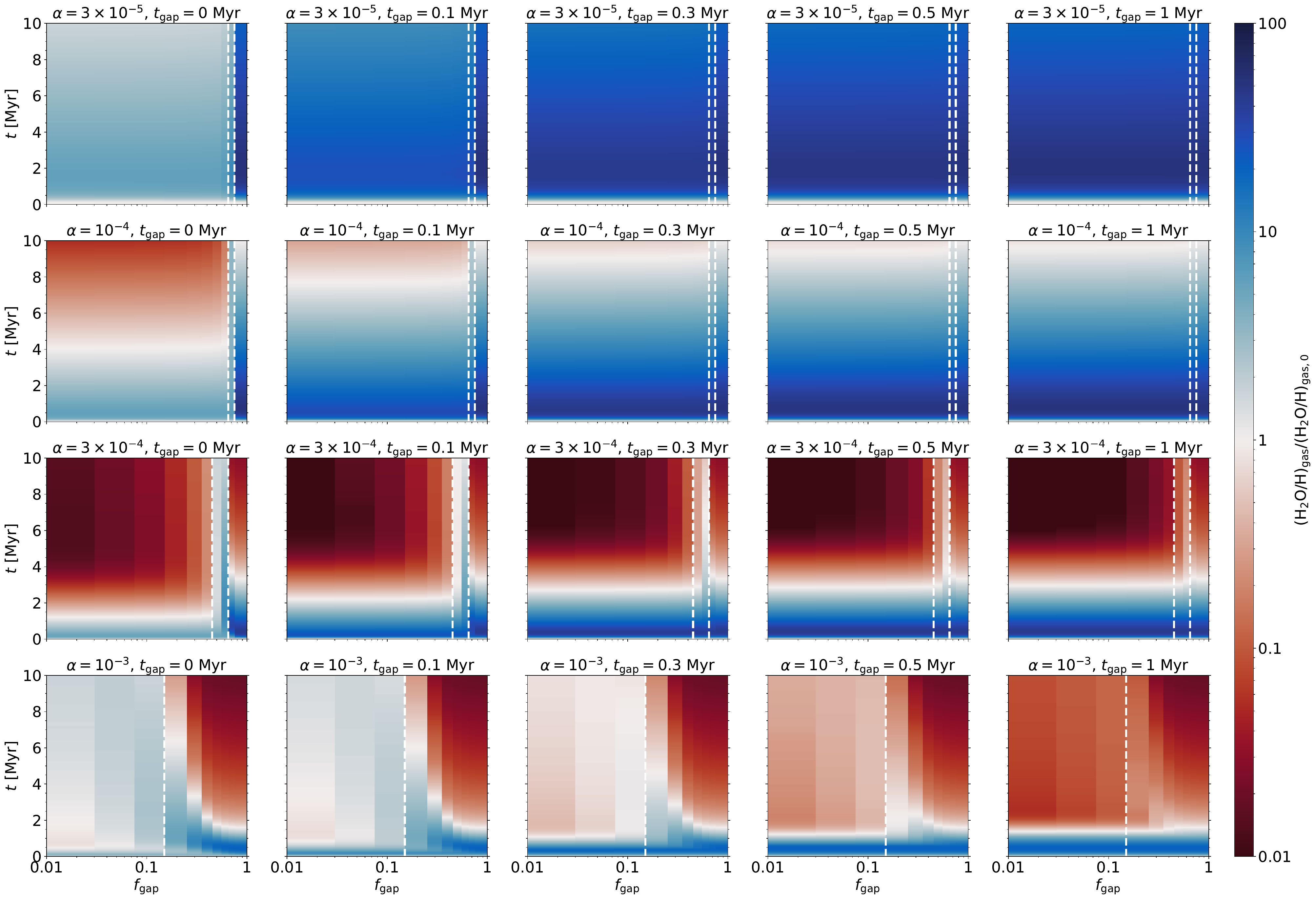}}
   \caption{Time evolution of the water vapour abundance at $r = 0.5~{\rm AU}$ normalised to the gas disc's initial value at this location ((H$_2$O/H)$_{\rm gas,0} = 3.54\times10^{-4}$), as a function of disc viscosity and the time of gap formation, where the insertion time of the gap increases from left to right. The gap is located at 10~AU.}
   \label{fig:bump-10au_water}
\end{figure*}

\begin{figure*}
\centering
   \resizebox{\hsize}{!}{\includegraphics{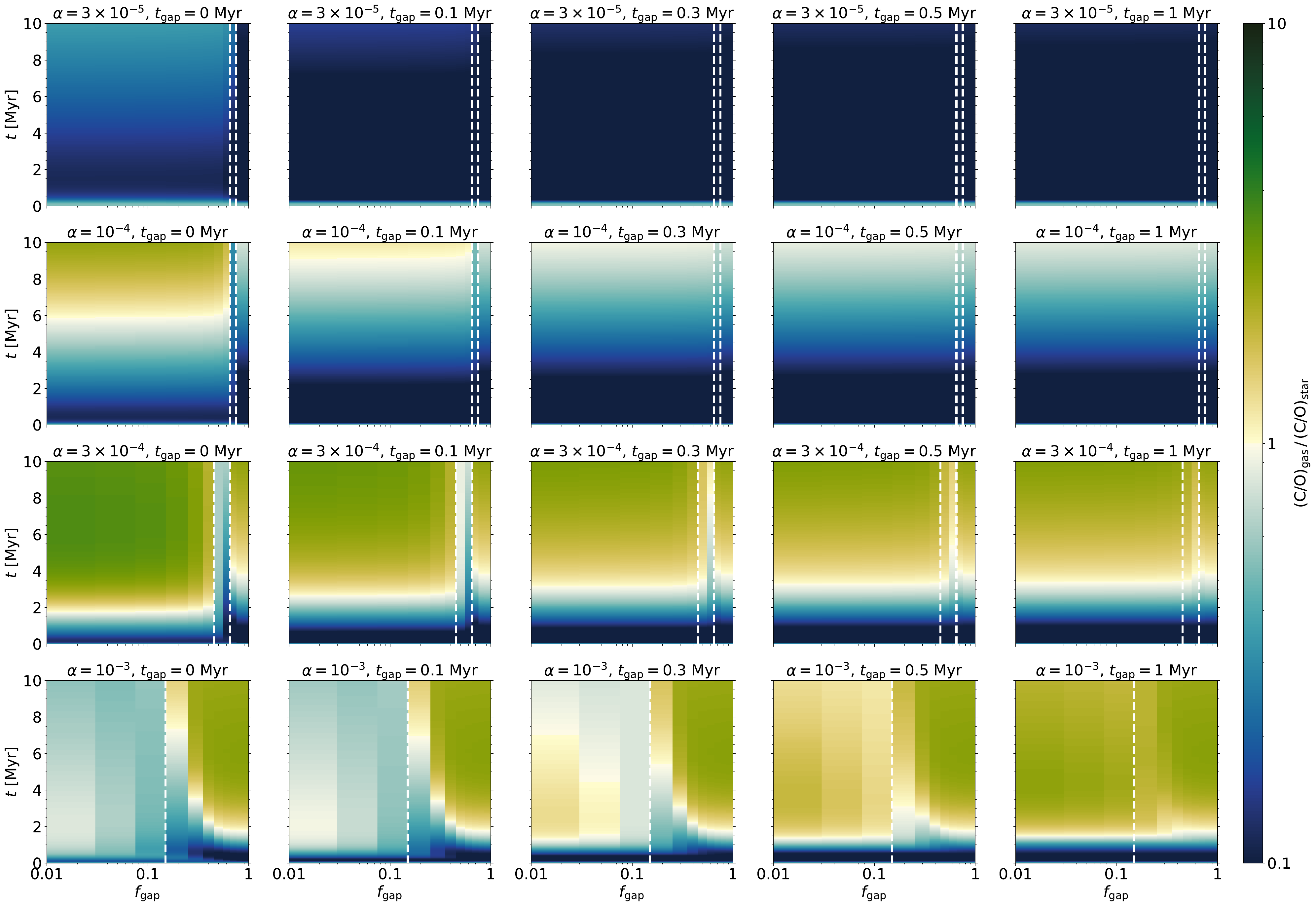}}
   \caption{Similar to Fig.~\ref{fig:bump-10au_water} but showing the time evolution of the C/O ratio at $r = 0.5~{\rm AU}$ normalised to the stellar value ((C/O)$_{\rm star} = 0.55$).}
   \label{fig:bump-10au_CO}
\end{figure*}

\begin{figure*}
\centering
   \resizebox{\hsize}{!}{\includegraphics{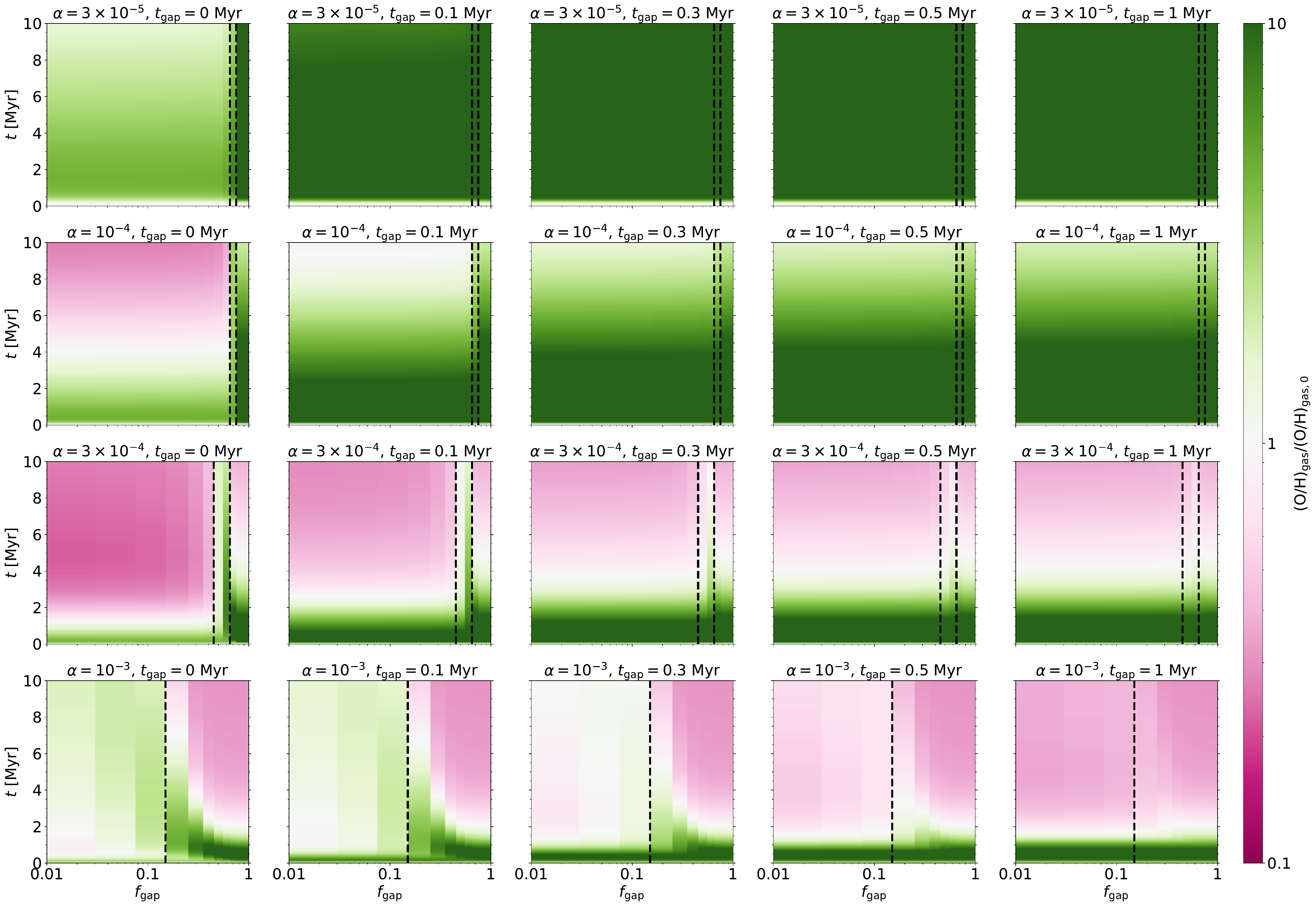}}
   \caption{Time evolution of the oxygen abundance at $r = 0.5~{\rm AU}$, normalised to the disc's initial value at this location ((O/H)$_{\rm gas,0} = 4.91\times10^{-4}$), as a function of disc viscosity $\alpha$ and the time when the gap forms $t_{\rm gap}$. The gap is located at 10~AU.}
   \label{fig:bump-10au_oxygen}
\end{figure*}

\begin{figure*}
\centering
   \resizebox{\hsize}{!}{\includegraphics{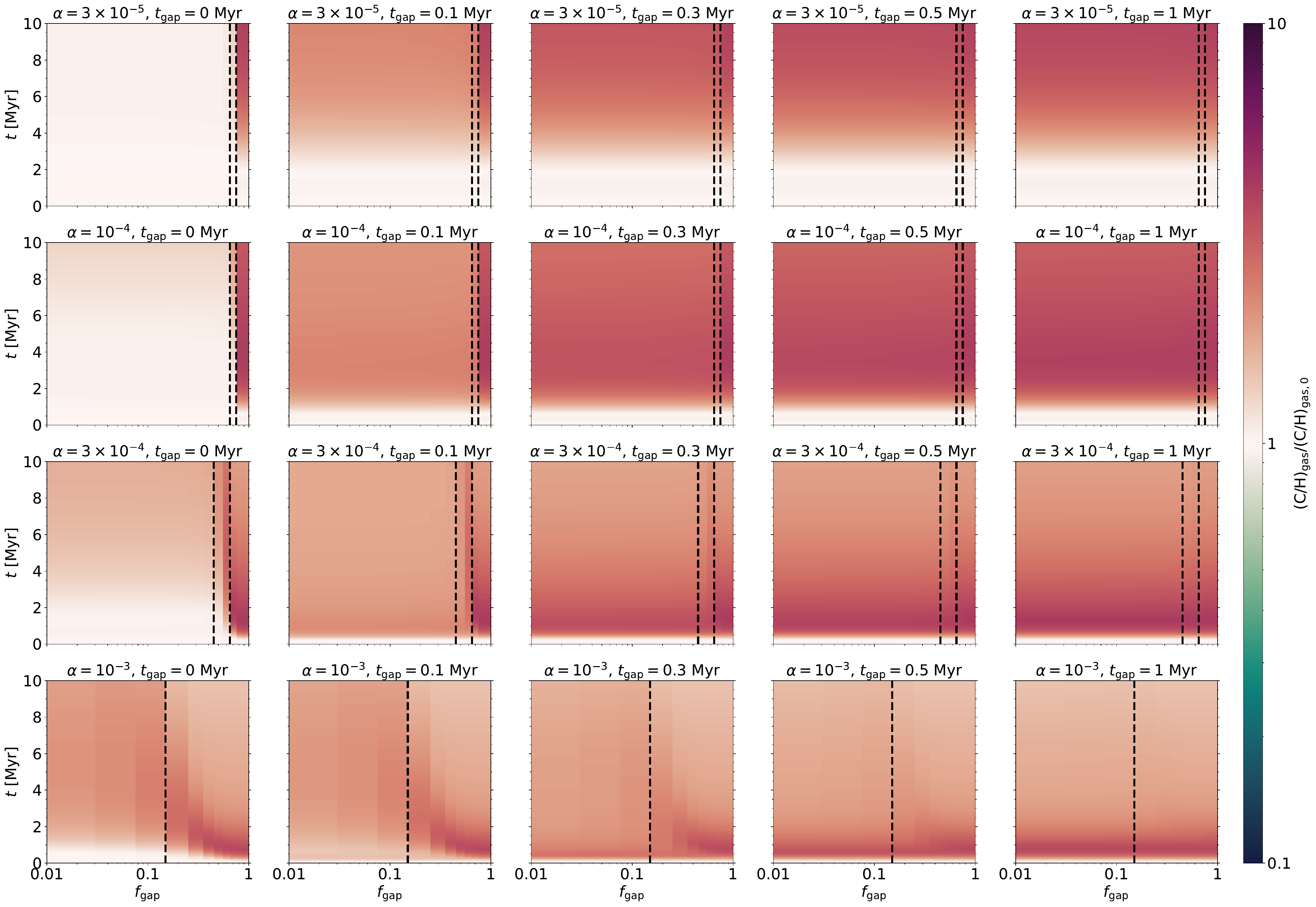}}
   \caption{Same as Fig.~\ref{fig:bump-10au_oxygen} but for carbon. The disc's initial carbon abundance at $r=0.5~{\rm AU}$ is (C/H)$_{\rm gas,0} = 1.37\times10^{-4}$. }
   \label{fig:bump-10au_carbon}
\end{figure*}

Here, we show the plots illustrating how the gap formation time influences the time evolution of the water abundance (Fig.~\ref{fig:bump-10au_water}), the C/O ratio (Fig.~\ref{fig:bump-10au_CO}), the O/H ratio (Fig.~\ref{fig:bump-10au_oxygen}), and the C/H ratio (Fig.~\ref{fig:bump-10au_carbon}) in the inner disc.

\section{Connecting gap depth to planet mass}
\label{appendix:fgap_vs_q}
Assuming that the gap is created by a growing planets, we can work out the gap depth as a function of planet mass following the prescription of \citet{Cridaetal2006} and \citet{CridaMorbidelli2007}. The gap depth is given by \citep{CridaMorbidelli2007}
\begin{equation}
\label{eq:f_p}
    f(\mathcal{P}) = 
        \begin{cases}
            (\mathcal{P}-0.541)/4               &\text{if $\mathcal{P} < 2.4646$,}\\
            1 - \exp{[-(\mathcal{P}^{0.75}/3)]} &\text{if $\mathcal{P} \geq 2.4646$,}
        \end{cases}
\end{equation}
where $\mathcal{P}$ is related to the properties of the planet and the disc following \citep{Cridaetal2006}:
\begin{equation}
    \mathcal{P} = \frac{3}{4}\frac{H_{\rm g}}{R_{\rm H}}+\frac{50}{q\mathcal{R}},
\end{equation}
where $H_{\rm g}$ is the gas scale height, $R_{\rm H} = a_{\rm pl}(M_{\rm pl}/3M_*)^{1/3}$ is the Hill radius of the planet, $a_{\rm pl}$ is the semi-major axis of the planet, $q = M_{\rm pl}/M_*$ is the planet-to-star mass ratio, and $\mathcal{R} = a_{\rm pl}^2\Omega_{\rm K}\nu$ is the Reynolds number. In our model, we use $f_{\rm gap} = f(\mathcal{P})$.

\begin{figure*}
\centering
   \resizebox{\hsize}{!}{\includegraphics{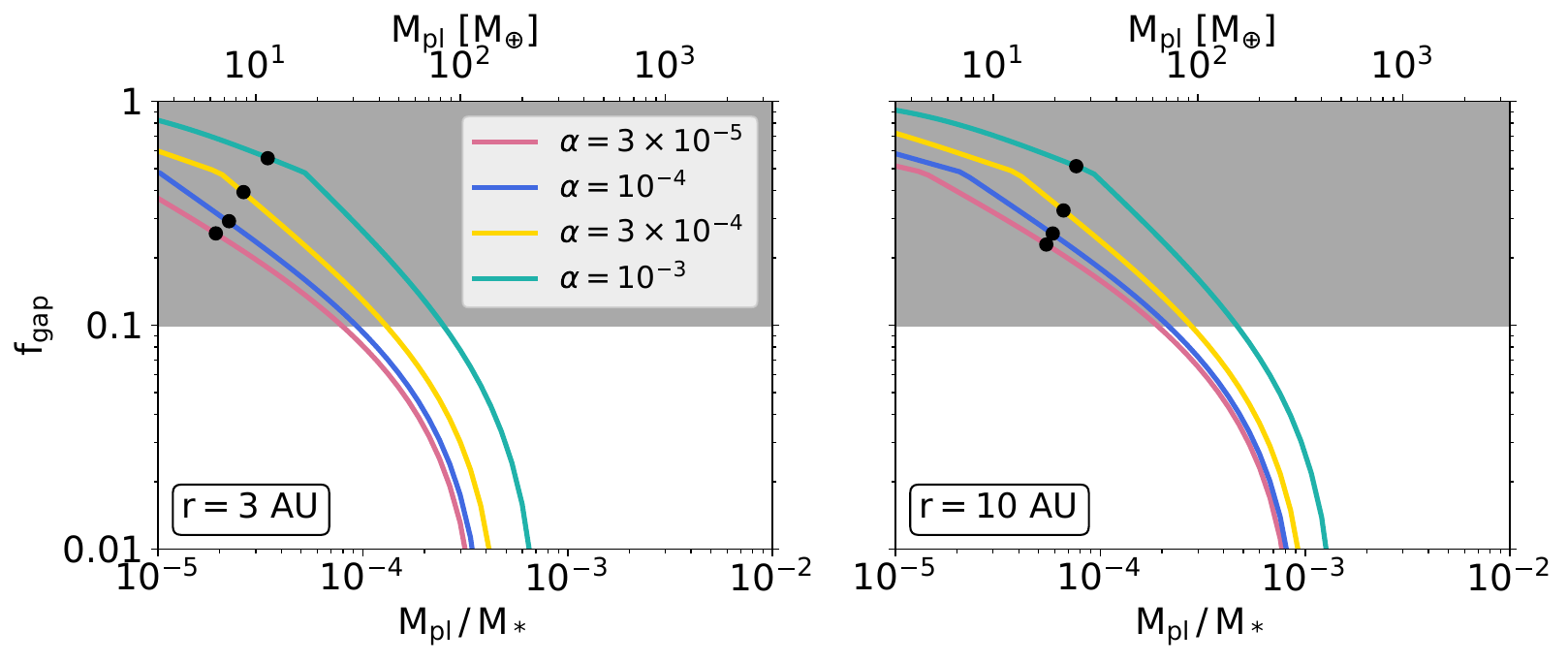}}
   \caption{Gap depth as a function of planet mass for different disc viscosities at $r = 3~{\rm AU}$ and 10~AU. Grey bands mark the range of values for $f_{\rm gap}$ that can be generated by zonal flows \citep{Flocketal2015}. Black dots mark the pebble isolation masses as computed from \citet{Bitschetal2018}. For $\alpha \leq 3\times10^{-4}$, the pebble isolation masses are in the deep gap regime while higher $\alpha$ results in pebble isolation masses within the shallow gap regime.}
   \label{fig:fgap_vs_q}
\end{figure*}

In Fig.~\ref{fig:fgap_vs_q}, we show how Eq.~\ref{eq:f_p} changes with $q$ at $r = 3~{\rm AU}$ and 10~AU for different disc viscosities. The gap depth increases ($f_{\rm gap}$ decreases) with increasing planet mass, as expected. More importantly, the plot also shows that we need more massive planets to achieve the same value of gap depth when: (1) the disc viscosity increases or (2) the planet is located further out in the disc. This is related to the gas scale height which increases with disc viscosity and orbital distance \citep{Bitschetal2015,Savvidouetal2020}. 

Our simulations are also consistent with the concept of the pebble isolation mass \citep[e.g.][]{MorbidelliNesvorny2012,Lambrechtsetal2014,Ataieeetal2018,Bitschetal2018}, namely planets that have reached the isolation mass cause deep gaps in the disc that block inward drifting pebbles rather than traffic jams. On the other hand, traffic jams could be caused by planets that are still growing by pebble accretion and do not open deep enough gaps yet.

\end{appendix}
\end{document}